# Spin-orbit coupling suppression and singlet-state blocking of spin-triplet Cooper pairs

Sachio Komori[1]*, James M. Devine-Stoneman[1], Kohei Ohnishi[2,3], Guang Yang[1], Zhanna Devizorova[4,5], Sergey Mironov[6], Xavier Montiel[1], Linde A. B. Olde Olthof[1], Lesley F. Cohen[7], Hidekazu Kurebayashi[8], Mark G. Blamire[1], Alexandre I. Buzdin[9,10], Jason W. A. Robinson[1]*

*sk891@cam.ac.uk      *jjr33@cam.ac.uk

[1] Department of Materials Science & Metallurgy, University of Cambridge, 27 Charles Babbage Road, Cambridge CB3 0FS, United Kingdom

[2] Department of Physics, Kyushu University, 744 Motooka, Fukuoka 819-0395, Japan

[3] Research Center for Quantum Nano-Spin Sciences, 744 Motooka, Fukuoka, 819-0395, Japan

[4] Moscow Institute of Physics and Technology, 141700 Dolgoprudny, Russia

[5] Kotelnikov Institute of Radio-engineering and Electronics RAS, 125009 Moscow, Russia

[6] Institute for Physics of Microstructures, Russian Academy of Sciences, 603950 Nizhny Novgorod, GSP-105, Russia

[7] The Blackett Laboratory, Imperial College London, SW7 2AZ, United Kingdom

[8] London Centre for Nanotechnology and Department of Electronic and Electrical Engineering at University of College London, London WC1H 01H, United Kingdom

[9] University Bordeaux, LOMA UMR-CNRS 5798, F-33405 Talence Cedex, France

[10] Sechenov First Moscow State Medical University, Moscow, 119991, Russia

**An inhomogeneous magnetic exchange field at a superconductor/ferromagnet interface converts spin-singlet Cooper pairs to a spin-aligned (i.e. spin-polarized) triplet state. Although the decay envelope of such triplet pairs within ferromagnetic materials is well studied, little is known about their decay in non-magnetic metals and superconductors, and in particular in the presence of spin-orbit coupling (SOC). Here we investigate devices in which triplet supercurrents are injected into the *s*-wave superconductor Nb. In the normal state of Nb, triplet supercurrents decay over a distance of 5 nm, which is an order of magnitude smaller than the decay of spin singlet pairs due to the SOC interacting with the spin associated with triplet pairs. In the superconducting state of Nb, triplet supercurrents are not able to couple with the singlet wavefunction and thus blocked by the absence of available equilibrium states in the singlet gap. The results offer new insight into the dynamics between *s*-wave singlet and *s*-wave triplet states.**

## INTRODUCTION

Spin-information can be transferred between ferromagnets through a superconducting spacer via spin-polarized quasiparticles or spin-polarized triplet Cooper pairs. Below the critical temperature of an *s*-wave superconductor, an energy gap opens in the density of states below which the electrons form pairs with antiparallel spins in a singlet state meaning singlet supercurrents do not carry a net spin. However, in this state the spin-relaxation time for spin-polarized quasiparticle (i.e. non-superconducting carrier) currents injected from a ferromagnet into a superconductor at the energy gap edge, is enhanced by 6 orders of magnitude over the normal state (*1*, *2*). Spin can also be carried directly in the superconducting state through the conversion of singlet pairs into spin-polarized triplet pairs (*3–5*) at magnetically inhomogeneous superconductor/ferromagnet (S/F) interfaces via spin-mixing and spin-rotation processes (*3–10*). Spin-triplet Cooper pairs have a spin degree of freedom and triplet supercurrents carry a net spin-polarization. For *s*-wave spin-triplet pairs, the antisymmetric wavefunction under an overall exchange of fermions is maintained through the odd-frequency pairing state (*11*, *12*). The majority of experiments to detect triplet supercurrents are based on S/$F_L$/F/$F_R$/S devices (*9*) in which the magnetization directions of the $F_L$ and $F_R$ layers are non-collinear to the magnetization direction of the central F. Examples include Nb/Ni/Cu/Co/Ru/Co/Cu/Ni/Nb devices (*13*, *14*) in which the magnetization directions of the outer Ni layers are orthogonal to the magnetization of the Co/Ru/Co synthetic antiferromagnet and Nb/Cr/Fe/Cr/Nb devices (*15*) where a spin-glass layer at the Fe/Cr interface provides magnetic inhomogeneity (*15–18*). Recently, ferromagnetic resonance spin-pumping experiments in Pt/Nb/Py/Nb/Pt structures have shown evidence for superconducting pure spin currents. In these structures the strong spin-orbit coupling (SOC) in Pt in conjunction with a proximity-induced ferromagnetic exchange field from Py creates a triplet density of states in superconducting Nb through which pure spin currents pumped from Py can propagate with a greater efficiency than when Nb is in the normal state (*19*, *20*).

Triplet pairs offer the potential for controlling spin and charge degrees of freedom via superconducting phase coherence (*3*, *4*, *21–23*); however, triplet device development requires an understanding of the decay envelope of generated triplet pairs in F, S and N (nonmagnetic) metals (i.e. the coherence length of triplet pairs extracted from the source S), as well as an understanding of the dynamic interaction of singlet and triplet states.

Spin-mixing and spin-rotation at an interface or a magnetic exchange field with SOC (*19*, *20*, *24*, *25*) are required to transform singlet pairs into triplet pairs. Away from such an interface the triplet pairs that are already formed should propagate through a second interface into an F, N, or S metal and transfer spin and the triplet wavefunction through these layers. In a ferromagnet, triplet pairs remain coherent over of tens of nanometers (*13–15*, *26*) and potentially hundreds of nanometers in half-metallic ferromagnets (*27*, *28*). Although little work has been done to explore triplet decay lengths in N metals, it is assumed that triplet pairs will remain coherent in N over the spin-diffusion length (*6*, *13*). Hence, a significant difference in the proximity decay lengths of singlet and triplet pairs is expected in N metals since SOC will scatter the net spin carried by a triplet supercurrent (*6*, *25*) and not the charge carried by a zero net spin singlet supercurrent.

A significant difference in the decay lengths is also expected for triplet and singlet pairs within an *s*-wave S. An attraction between electrons with opposite spin projections inside the *s*-wave superconductor supports the transfer of singlet pairs through the S layer without any damping. However, the triplet pairs that penetrate a superconductor experience the spatial decay of their wavefunction since the singlet gap does not support electrons with equal spin projections.

Here we investigate the triplet coherence in Nb, a metal with strong SOC (*29–31*). The triplet coherence length is investigated in both the normal and superconducting states by fabricating four series of S/$F_L$/S'/$F_R$/S devices: (A) "triplet control devices" Nb(300)/Cr(1)/Fe($d_{Fe}$)/Cr(1)/Nb(300) (thicknesses in nm units) without S' (also denoted Nb') and varying the total thickness of Fe from 3 to 15 nm to confirm singlet-to-triplet pair conversion at the Cr/Fe and Fe/Cr spin-mixer interfaces; (B) "singlet devices" Nb(300)/Cr(1)/Fe(2)/Nb'($d_{Nb'}$)/Fe(2)/Cr(1)/Nb(300) in which the total Fe thickness is low enough such that a residual singlet supercurrent is measurable; and two series of "triplet devices" with (C) Nb(300)/Cr(1)/Fe(4.8)/Nb'($d_{Nb'}$)/Fe(2.4)/Cr(1)/Nb(300) and
(D) Nb(300)/Cr(1)/Fe(7.5)/Nb'($d_{Nb'}$)/Fe(2.0)/Cr(1)/Nb(300) layers with a total Fe thickness exceeding the maximum thickness for which a singlet supercurrent is observed in Nb/Fe/Nb devices (5.5 nm) (*15*). Each set of devices were prepared in a single deposition run. In device series (B) – (D), there are no intentional spin-mixing and spin-rotation interfaces between the Fe layers and the central Nb' layer and hence a triplet pair wavefunction should not be generated in Nb' in the superconducting state.

Current-perpendicular-to-plane S/$F_L$/S'/$F_R$/S Josephson devices are fabricated using a focused ion beam microscope technique that is described in detail elsewhere (*32*). Due to variations in the cross-sectional areas of the devices, the Josephson critical current ($I_c$) is multiplied by the device normal state resistance $R_n$ (estimated from the voltage at high current bias) to give the characteristic voltage ($I_cR_n$). The $I_cR_n$ of all devices is systematically investigated as a function of $d_{Nb'}$ in the 0 to 40 nm range.

**RESULTS AND DISCUSSION**

We first discuss the triplet control devices. In Fig. 1A, we compare $I_cR_n$ versus Fe layer thickness ($d_{Fe}$) of these devices with the known $d_{Fe}$-decay envelope of $I_cR_n$ for Nb/Fe/Nb (blue curve) and Nb/Fe/Cr/Fe/Nb (black curve) devices previously measured by our group (*15*). The Nb/Fe/Nb and Nb/Fe/Cr/Fe/Nb devices do not have Nb/Cr/Fe (Fe/Cr/Nb) spin-mixer interfaces and so transport is spin-singlet. The Nb/Fe/Cr/Fe/Nb devices have the same number of interfaces as the Nb/Cr/Fe/Cr/Nb triplet devices and therefore acts as better control devices than the Nb/Fe/Nb devices. For $d_{Fe}$ < 5 nm, supercurrents are detectable in both triplet and singlet devices, but for $d_{Fe}$ > 5 nm supercurrents are only detectable in the triplet control devices confirming spin-mixing and spin-rotation at the Nb/Cr/Fe (Fe/Cr/Nb) interfaces. The deviation from the exponential fit for the device with $d_{Fe}$ = 6 nm is probably due to the sample-to-sample variation.

By applying a magnetic field ($H$) parallel to the interfaces, the $I_c$ of the triplet control devices is modulated (inset of Fig. 1B). $I_c(H)$ is hysteretic and the maximum values of $I_c$ are obtained at non-zero applied field ($\mu_0 H = \delta$) due to the barrier magnetization. In Fig. 1B we have plotted $\delta$ at 1.6 K (left-axis) versus $d_{Fe}$, which shows a linear increase in $\delta$ with $d_{Fe}$, consistent with the linear rise in the magnetic moment ($m_s$) per unit area with $d_{Fe}$ for the unpatterned Nb/Cr/Fe/Cr/Nb films measured using a vibrating sample magnetometer at 300 K (right-axis). Both $\delta$ and $m_s$ per unit area are proportional to $d_{Fe}$, suggesting that the Fe layers are homogeneously magnetized at magnetic saturation in both the unpatterned films and devices. From Fig. 1B, we estimate a magnetically dead layer at each Fe/Cr interfaces of 0.2 – 0.3 nm, which likely constitutes a spin-glass (*15–18*).

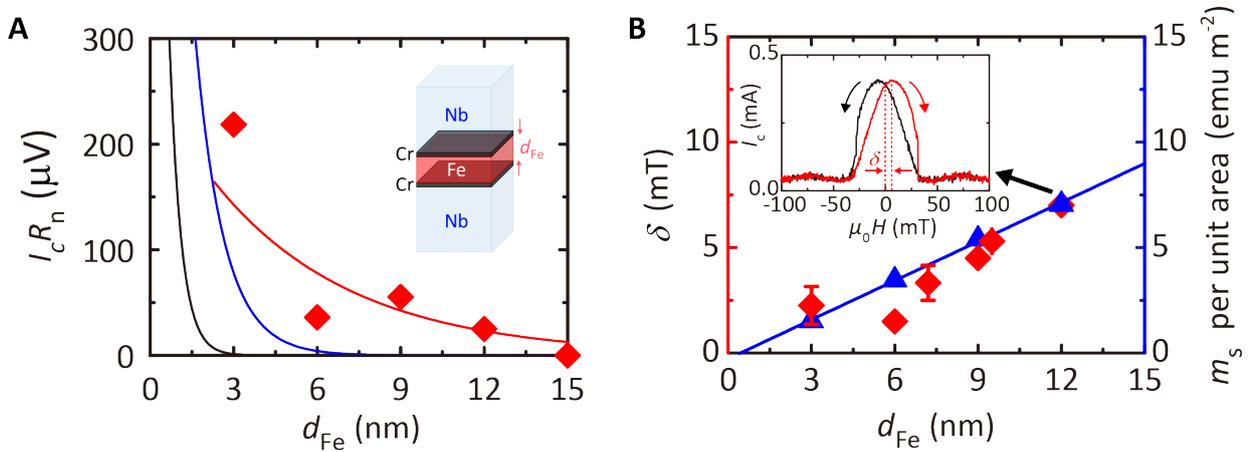

**Fig. 1. Triplet pair creation and magnetization properties of Fe/Cr interfaces.** (**A**) The decay of the average critical current multiplied by the normal state resistance ($I_c R_n$) versus Fe thickness ($d_{Fe}$) for Nb(300)/Cr(1)/Fe($d_{Fe}$)/Cr(1)/Nb(300) triplet control devices (red diamonds) at 1.6 K along with the known $d_{Fe}$-decay of $I_c R_n$ for Nb(300)/Fe($d_{Fe}$)/Nb(300) (the blue curve) singlet devices with a coherence length of 1.0 nm (*15*) and Nb(300)/Fe($d_{Fe}$/2)/Cr(2.5)/Fe($d_{Fe}$/2)/Nb(300) (the black curve) singlet devices with a coherence length of 0.5 nm (*15*). The (red) curve is a least square fit giving a triplet coherence length of $\xi_F^{triplet}$ = 5.3 ± 1.9 nm. (**B**) In-plane magnetic hysteresis ($\delta$; red diamonds, left axis) estimated from the Nb(300)/Cr(1)/Fe($d_{Fe}$)/Cr(1)/Nb(300) triplet control devices at 1.6 K where $\delta$ is the maximum field shift in $I_c(H)$. The right axis shows the magnetic moment at magnetic saturation per unit area ($m_s/m^2$) determined from unpatterned Nb(300)/Cr(1)/Fe($d_{Fe}$)/Cr(1)/Nb(300) films (blue triangles). The (blue) curve is a least-squares regression line fit to $m_s/m^2$ versus $d_{Fe}$ with a volume magnetization of 618 emu cm$^{-3}$ and a magnetically dead layer at each Fe/Cr interfaces of 0.2 – 0.3 nm. The inset shows an $I_c(H)$ pattern for a Nb(300)/Cr(1)/Fe(12)/Cr(1)/Nb(300) device at 1.6 K.

In Fig. 2A we have plotted $I_cR_n$ versus $d_{Nb'}$ for the singlet devices which show two Nb'-thickness regimes: for $d_{Nb'}$ < 30 nm, $I_cR_n$ slowly decreases with increasing $d_{Nb'}$ and rises beyond 30 nm, indicating the onset of superconductivity in Nb' which leads to two Josephson devices operating in series with the effective barrier thickness reduced as illustrated in Figs. 2B and 2C. Since the potential injection of spin-polarized quasiparticles suppresses the onset superconductivity of Nb', it is difficult to distinguish the critical current of Nb' and the Josephson critical current of the two devices. However, the formation of the two Josephson devices in series is confirmed by a second harmonic Fraunhofer pattern which results from the overlap of the Andreev bound states in Nb'(*33–35*). In Fig. 2D, we have plotted the positive field direction in $I_c$ (*H*) for two representative devices for two different values of $d_{Nb'}$ (20 and 30 nm). $I_c$ is modulated with magnetic flux [$\Phi = \mu_0HL(2\lambda+d)$] according to sinc ($n\Phi/\Phi_0$), but the periodicity (1/*n*) is halved (*n* = 2) for the 30 nm device, consistent with a second harmonic current-phase relationship. Here, *L* is the length of the junction perpendicular to the field, $\lambda$ = 110 nm (*36*, *37*) is an estimate of the London penetration depth of Nb, *d* is the effective barrier thickness and $\Phi_0$ is a flux quantum. In Fig. 2E (left-axis), we have plotted *n* versus $d_{Nb'}$, which shows *n* = 1 behaviour for all thicknesses except for the 30 nm device (which matches the singlet coherence length). The *n* = 1 behaviour (i.e. the first harmonic) for the $d_{Nb'}$ = 40 nm devices is consistent with weakly overlapped Andreev bound states (*33*, *34*). To calculate *n*, we used $d = d_{Nb'} + 2d_{Cr}$ (2 nm) $+ 2d_{Fe}$ (4 nm) for $d_{Nb'}$ < 30 nm and $d = d_{Cr}$ (1 nm) $+ d_{Fe}$ (2 nm) for $d_{Nb'} \geq$ 30 nm. The relatively large error of *n* for $d_{Nb'}$ = 30 nm indicates the crossover between the conventional first harmonic and the unconventional second harmonic behaviour. The non-zero $I_c$ (*H*) minima for the $d_{Nb'}$ = 30 nm device may be due to the non-uniform supercurrent mediated by the superconducting Nb'.

From the total specific resistance of these devices ($AR_n$) versus $d_{Nb'}$ (Fig. 2E; right-axis) and fitting a least-squares regression line, we estimate a resistivity in Nb' of $\rho_{Nb'} \approx 7.8 \pm 1.1$ $\mu\Omega$·cm (where *A* is the device cross-sectional area). The effective electron mean free path is $l = m_e v_F / N_d \rho_{Nb} e^2 \approx 11.2 \pm 1.4$ nm, where $m_e \approx 9.1\times10^{-31}$ kg is the (effective) electron mass, $v_F = 1.37 \times 10^6$ m s$^{-1}$ is the Fermi velocity of Nb (*38*), $N_d = 5.56\times10^{28}$ m$^{-3}$ is the number density of conduction electrons in Nb (*38*), and *e* is the electric charge. The electron diffusivity is $D_N = v_F l/3 \approx (5.1 \pm 0.6) \times 10^{-3}$ m$^2$ s$^{-1}$, which gives a singlet coherence length of $\xi_N^{singlet} = (\hbar D_N/2\pi k_B T)^{1/2} \approx 61 \pm 4$ nm, consistent with the decay of $I_cR_n$ versus $d_{Nb'}$ for $d_{Nb'}$ < 30 nm.

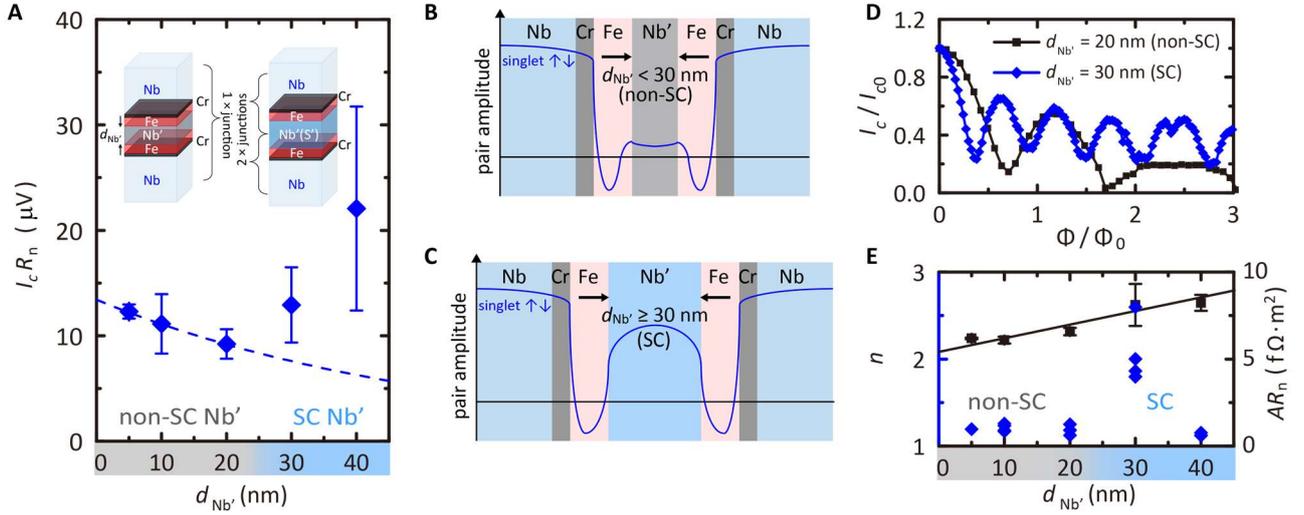

**Fig. 2. Supercurrents in singlet devices.** (**A**) Characteristic voltage ($I_cR_n$) versus $d_{Nb'}$ at 1.6 K for Nb(300)/Cr(1)/Fe(2)/Nb'($d_{Nb'}$)/Fe(2)/Cr(1)/Nb(300) devices. The dashed curve is a least square regression line fit to $I_cR_n$ for $d_{Nb'}$ < 25 nm, giving a singlet coherence length of $\xi_N^{singlet}$ = 52 ± 22 nm. The error bars in $I_cR_n$ represent the statistical average values of $I_cR_n$ for multiple devices at a given value of $d_{Nb'}$, taking into account the errors in $I_c$ and $R_n$. (**B** and **C**) Schematic illustrations of the superconducting pair amplitudes for $d_{Nb'}$ < 30 nm and $d_{Nb'}$ ≥ 30 nm. (**D**) Normalized critical current versus normalized magnetic flux ($\Phi/\Phi_0$) for Nb(300)/Cr(1)/Fe(2)/Nb'(20)/Fe(2)/Cr(1)/Nb(300) (black squares) and Nb(300)/Cr(1)/Fe(2)/Nb'(30)/Fe(2)/Cr(1)/Nb(300) (blue diamonds) devices at 1.6 K. (**E**) Normalized inverse periodicity of Fraunhofer oscillation (*n*; blue diamonds, left-axis) and specific normal state resistance ($AR_n$; black squares, right-axis) versus $d_{Nb'}$ for the singlet devices at 1.6 K with a least-squares regression line fit to $AR_n$ (black line) from which we estimate $\rho_{Nb'}$ ≈ 7.8 ± 1.1 $\mu\Omega\cdot$cm. $R_n$ values for $d_{Nb'}$ = 30 and 40 nm are taken when Nb' is in the normal state.

The trend in $I_cR_n$ versus $d_{Nb'}$ for both sets of triplet devices at 1.6 K is different to the singlet devices in that they do not show two-series junction behaviour for all values of $d_{Nb'}$ investigated (see Fig. 3). For $d_{Nb'}$ < 15 nm, Josephson coupling is achieved (see Fig. 3B) and the corresponding normal-state resistance of the devices (*R*) falls to zero below 4 K. The inset of Fig. 3A shows an $I_c(H)$ pattern for a Nb(300)/Cr(1)/Fe(7.5)/Nb'(4)/Fe(2.0)/Cr(1)/Nb(300) device showing standard Fraunhofer behaviour with $I_c$ shifted in field due to barrier flux from Fe. The periodicity of the Fraunhofer oscillation in the triplet devices is 77 – 86% of the first harmonic (*n* = 1) Fraunhofer pattern for a magnetized junction (*39*) (See top right of the inset in Fig. 3A). The slightly reduced *n* values and the slow decay of Fraunhofer oscillation may be due to the variation of the pair conversion efficiency within the device area.

Typical *R* (*T*) curves for $d_{Nb'}$ < 15 nm are shown in Fig. 3C. The 300-nm-thick top and bottom Nb layers become superconducting below 9 K, showing a drop in *R* with the resistance continuously decreasing with decreasing temperature as superconductivity gradually proximitizes the Cr/Fe/Nb'/Fe/Cr barrier. The barriers are completely proximitized (*R* = 0) below 4 K. The decay in $I_cR_n$ versus $d_{Nb'}$ is exponential [$I_cR_n = \exp(-\xi_N^{triplet}/d_{Nb'})$] with a triplet coherence length of ≈ 3.2 – 5.7 nm, which is an order of magnitude smaller than the singlet coherence length in Nb' estimated from Fig. 2A. The strong pair breaking effect on triplet pairs is likely due to strong SOC in normal state Nb (*29–31*), which suppresses the triplet pairing coherence due to scattering of the spin associated with the triplet supercurrent (*6, 25*). We note that, for all temperatures, we do not observe magnetoresistance from the Fe/Nb'/Fe barriers in these devices, suggesting a short spin-diffusion length in thin Nb' layers ( < 10 nm) in these particular devices due to SOC (*29–31*) (see Supplementary Materials for details).

In the $d_{Nb'}$ range 15 to 30 nm, $R$ of the devices does not fall to zero (Fig. 3E) and Josephson coupling is not detected (i.e. no $I_c$), suggesting the absence of triplet or singlet supercurrents, i.e. the triplet pair amplitude across Nb' is (approximately) zero. For $d_{Nb'} \geq 30$ nm, the Nb' spacers show a superconducting transition with dips in $R$ below 2.3 K and 5.0 K for $d_{Nb'}$ = 30 and 40 nm, respectively (Fig. 3G). The resistivity of the Nb' layer calculated from the resistance drop associated with the superconducting transition for these devices is 8.2 – 10.4 $\mu\Omega \cdot$cm, consistent with the value estimated from Fig. 2E. In contrast to the singlet devices (Fig. 2A) we do not observe two-series junction behaviour in which the superconducting Nb' layer effectively halves the thickness of the barrier layers and leads to a higher $I_c R_n$ over the normal state Nb', meaning that the triplet wavefunction is unable to mediate Josephson coupling with the singlet wavefunction of Nb'. The triplet supercurrent is blocked even for the device with the thinnest superconducting Nb' layer ($d_{Nb'}$ = 30 nm) obtained in this work and hence we estimate the coherence length of triplet pairs to be shorter than the singlet pair correlation length ( $\approx$ 30 nm). The disconnection of the triplet pair amplitude across the Nb' layer blocks charge transport via triplet pairs, i.e. Nb' is an effective insulator for triplet pairs.

In a related experiment, we investigated the superconducting density of states (DoS) on NbN/La$_{2/3}$Ca$_{1/3}$MnO$_3$ using scanning tunneling microscopy (*40*), where NbN is an *s*-wave superconductor and La$_{2/3}$Ca$_{1/3}$MnO$_3$ is a highly spin-polarized ferromagnetic manganite. Here an enhancement of the superconducting DoS in NbN was observed around zero energy, consistent with spin-one triplet theory assuming a magnetically inhomogeneous interface (*41*). In agreement with the present manuscript, the zero energy enhancement of the DoS in NbN rapidly decayed as a function of NbN thickness with a decay envelope close to the spin-diffusion and superconducting coherence lengths; these results demonstrated that the proximity-induced triplet state in NbN was unfavourable within an intrinsic singlet DoS.

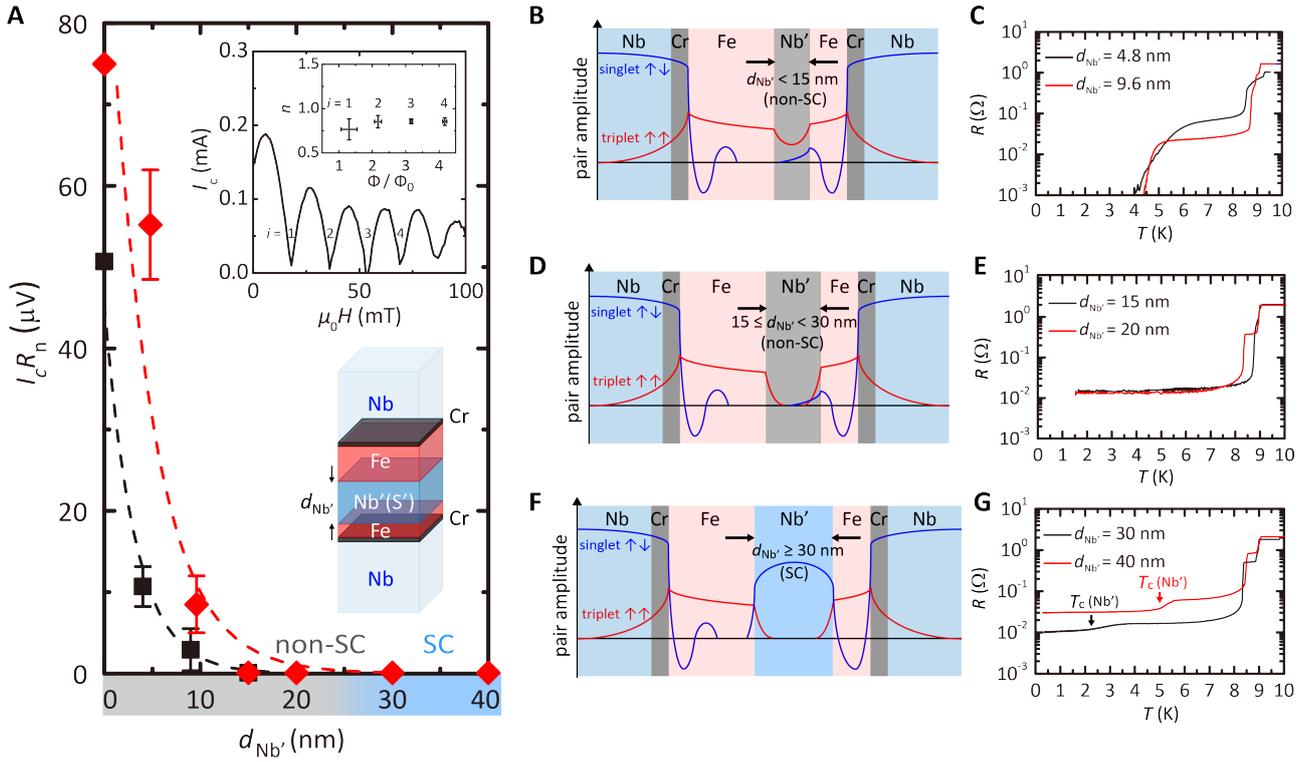

**Fig. 3. Supercurrents in triplet devices.** (**A**) Characteristic voltage $I_cR_n$ versus thickness of the central Nb' layer ($d_{Nb'}$) at 1.6 K for Nb(300)/Cr(1)/Fe(4.8)/Nb'($d_{Nb'}$)/Fe(2.4)/Cr(1)/Nb(300) (red diamonds) and Nb(300)/Cr(1)/Fe(7.5)/Nb'($d_{Nb'}$)/Fe(2.0)/Cr(1)/Nb(300) (black squares) devices. The (red and black) dashed curves represent a least square regression line fit to $I_cR_n$, giving triplet coherence lengths of $\xi_N^{triplet}$ = 5.7 and 3.2 nm, respectively. The $I_cR_n$ values at $d_{Nb'}$ = 0 nm are estimated from the gradient of $I_cR_n$ versus $d_{Fe}$ for the Nb(300)/Cr(1)/Fe($d_{Fe}$)/Cr(1)/Nb(300) triplet devices in Fig. 1A. The error bars in $I_cR_n$ represent the statistical average values of $I_cR_n$ for multiple devices at a given value of $d_{Nb'}$, taking into account the errors in $I_c$ and $R_n$. The inset shows an $I_c(H)$ pattern for a Nb(300)/Cr(1)/Fe(7.5)/Nb'(4)/Fe(2.0)/Cr(1)/Nb(300) device at 1.6 K. (Top right of the inset) Distribution of intensity minimas in $I_c(H)$ oscillations as a function of external magnetic flux for the triplet devices and corresponding normalized frequency ($n$) calculated by fitting $I_c(H)$ to the standard Fraunhofer pattern (*39*). (**B**-**G**) Schematic illustrations of the superconducting pair amplitudes at 1.6 K and $R(T)$ curves for $d_{Nb'}$ < 15 nm, 15 ≤ $d_{Nb'}$ < 30 nm, and $d_{Nb'}$ ≥ 30 nm regimes in Nb(300)/Cr(1)/Fe(4.8)/Nb'($d_{Nb'}$)/Fe(2.4)/Cr(1)/Nb(300) devices.

The significant differences in the coherence lengths of singlet and triplet pairs observed in F (Fe), N (normal state Nb'), and S' (superconducting Nb') are summarized in Table 1 together with the mean free path and the spin diffusion lengths. In F, the coherence length of triplet pairs is long-ranged and close to the spin-diffusion length, while singlet pairs affected by the exchange field are short-ranged (Fig. 1A). In N with strong SOC, the coherence length of the triplet pairs is short-ranged (Fig. 3A) due to the short spin-diffusion length (see Supplementary Materials) while singlet pairs are unaffected by SOC and are long-ranged (Fig. 2A). In S', singlet pairs couple with the singlet wavefunction of S' and creates two-series junction behaviour and hence singlet supercurrents do not show a decay (Fig. 2A). Triplet pairs however are not able to couple with the singlet wavefunction of S' and hence decay within the order of the singlet coherence length (30 nm; Fig. 3A).

**Table 1. Electron mean free path (*l*), spin diffusion length (*l*$_{sd}$) and coherence lengths (*ξ*) in Fe and Nb at 1.6 K.**

| length scale (nm) | Fe | Nb' (non-SC) | Nb' (SC) |
|---|---|---|---|
| $l$ | 10.4 Ref. (*42*) | 11.2 ± 1.4 | – |
| $l_{sd}$ | 8.5 ± 1.5 Ref. (*42*) | < 4.8 | – |
| $\xi^{singlet}$ | 1.0 Ref. (*15*) | 52 ± 22 | no decay |
| $\xi^{triplet}$ | 5.3 ± 1.9 | 4.5 ± 1.3 | < 30 |

Triplet pairs which are not able to couple with the singlet wavefunction can be blocked in the singlet superconducting Nb' through SOC or (and) a competition with the singlet pairing correlation. There is no existing theory to explain the effect of SOC on triplet pairs in a material with a singlet pairing correlation. Assuming that the singlet pairing correlation of Nb' does not affect the SOC scattering of triplet pairs and there is no interaction between the singlet and the triplet pairing states, the decay length of the triplet pairs in the superconducting Nb' is obtained from the equation (5.36) in Ref. (*6*):

$$\xi_S^{triplet} \approx \left\{2\left(\frac{4}{\tau_{SO}D_N}\right)\right\}^{-\frac{1}{2}} = \frac{1}{2}\left(\frac{l_{so}l}{6}\right)^{\frac{1}{2}} \approx 0.2\xi_N^{singlet}\left(\frac{l_{so}}{\xi_0}\right)^{\frac{1}{2}} \quad \text{if} \quad \frac{4\hbar}{\tau_{SO}} \gg k_BT_c, \qquad (1)$$

where $\xi_0 = 0.18\hbar v_F/k_BT_c$, $l_{so}$ and $\tau_{so}$ are the mean free length and the mean free time between the spin-orbit scattering events, respectively. A rough estimation $l_{so} \approx l_{sd} \approx 5$ nm, $\xi_0 \approx 30$ nm and $\xi_N^{singlet} \approx 52 \pm 22$ nm gives $\xi_S^{triplet} \approx 4.2 \pm 1.8$ nm, consistent with the experimental results showing a blocking of triplet supercurrents in a singlet superconducting Nb' ($\xi_S^{triplet} < 30$ nm) and matching with $\xi_N^{triplet} \approx 4.5 \pm 1.3$ nm estimated from Fig. 3A.

However, in the presence of the singlet pairing correlation, triplet pairs would no longer experience an effective field due to the SOC since the condensate requires a matching density of states for up and down spin electrons – hence superconductivity and a supercurrent is immune to the presence of SOC. If this is the case, the strong suppression of triplet pairs is dominated by a competition between the singlet and the triplet pairing states (*43*) resulting from the fact that they have an opposite influence on the electron density of states at the Fermi energy, i.e. the singlet pairing decreases it, while the triplet correlations lead to its increase. To show the effect of singlet pairing correlation on the decay of triplet pairs, we calculate the critical current density in a S$_L$/F$_L$/S'/F$_R$/S$_R$ device where S$_L$/F$_L$ and F$_R$/S$_R$ are spin-mixing/rotation interfaces and each layer is atomically thin. The central S' layer has a superconducting gap of $\Delta_0$ which is smaller than that of $S_L$ and $S_R$ ($\Delta_1$). The magnetic exchange fields of $F_L$ and $F_R$ layers (spin-rotation axis) are parallel to each other and strong enough to block the transport of minority spin triplet pairs. By solving the Gor'kov equations derived from a hopping probability of electrons between the atomically thin layers, (see Supplementary Materials for details), we obtain the critical current density which appears to be completely triplet:

$$J_c = |\Delta_1|^2 h_L h_R \sin\theta_L \sin\theta_R (a - b|\Delta_0|^2), \qquad (2)$$

where $h_L$ ($h_R$) is the magnetic exchange field in $F_L$ ($F_R$) and $\theta_L$ ($\theta_R$) is the magnetization angle between the magnetic exchange field at the $S_L/F_L$ ($F_R/S_R$) interface and $F_L$ ($F_R$). We note that equation (2) obtained from the anomalous Green's functions in S' consists of only triplet supercurrents and a singlet component is absent, meaning that phase-coupling between triplet pairs and the singlet wavefunction in S' is not mediated, agreeing with the experimental results. Since the coefficients $a$, $b > 0$, the presence of a singlet gap in S' layer ($\Delta_0$) suppresses the triplet current density. This results from the fact that $\Delta_0$ suppresses the triplet component of the anomalous Green's function (i.e. the motion of triplet pairs), which also agrees with the decay of triplet pairs within the length scale of singlet coherence length shown in Fig. 3.

**CONCLUSION**

We have observed a strong suppression of spin-triplet supercurrents in the normal and superconducting states of the *s*-wave superconductor Nb. In the normal state, SOC rapidly scatters triplet pairs and in the superconducting state triplet pairs are not able to mediate phase-coupling and are blocked, qualitatively consistent with our theoretical model. Although the exact underlying mechanism(s) for triplet pair suppression in an *s*-wave gap remains an open question, the results provide insight into the dynamic coupling of *s*-wave singlet and *s*-wave triplet states demonstrating a mechanism for superconducting filtering of triplet pairs.

**MATERIALS AND METHODS**

**Film growth**

Unpatterned films were fabricated on 5 mm × 5 mm quartz substrates by direct current magnetron sputtering in an ultrahigh-vacuum chamber with a base pressure better than $10^{-6}$ Pa. The sputtering targets were pre-sputtered for approximately 20 minutes to clean the surfaces and the films were grown using an Ar pressure of 1.5 Pa. Multiple quartz substrates were placed on a rotating circular table that passed in series under stationary magnetrons so that multiple samples with different layer thicknesses could be grown in the same deposition run. The thickness of each layer was controlled by adjusting the angular speed of the rotating table at which the substrates moved under the respective targets and the sputtering power.

**Device fabrication**

Standard optical lithography and Ar-ion milling define 4-μm-wide tracks, which were narrowed using a focused beam of Ga ions (Zeiss Crossbeam 540) to make current-perpendicular-to-plane devices. Further details on the device fabrication process are described elsewhere (*32*). A typical device dimension is 500 nm × 500 nm.

**Transport measurements**

A pulse-tube cryogen-free system (Cryogenic Ltd) was used to cool the devices down to 1.6 K. Resistivity and current-voltage $I(V)$ characteristics of the devices were measured in a four-point configuration using a current-bias circuit attached to a lock-in amplifier and an analogue-digital converter and also using the differential conductance mode of a Keithley 6221 AC-current source and a 2182A nanovoltmeter. The Josephson critical current $I_c$ and the normal state resistance $R_n$ of a device were determined by fitting the $I(V)$ characteristics to the resistively shunted junction model $V = R_n (I^2 - I_c^2)^{0.5}$.

**ACKNOWLEDGEMENTS**

**General:** The authors acknowledge K-R. Jeon for fruitful discussions.

**Funding:** S.K., J.M.D-S., G.Y., X.M., L.F.C., H.K., M.G.B. and J.W.A.R. acknowledge funding from the EPSRC Programme Grant "Superspin" (No. EP/N017242/1) and EPSRC International Network Grant "Oxide Superspin" (No. EP/P026311/1). K.O. acknowledges the JSPS Programme "Fostering Globally Talented Researchers" (JPMXS05R2900005). S.M. and A.I.B. acknowledge funding from Russian Science Foundation (Grant No. 20-12-00053, in part related to the theoretical calculations). Zh.D. and S.M. acknowledge financial support from the Foundation for the advancement of theoretical physics "BASIS". S.M. acknowledges financial support from the Russian Presidential Scholarship (SP-3938.2018.5).

**Author contributions:** J.W.A.R. conceived and designed the project. S.K. and J.M.D-S. sputtered the films, fabricated the devices and performed electrical and magnetic measurements. The electrical data was obtained using a setup developed by S.K. with help from K.O. and G.Y. The theoretical model was developed by Zh.D., S.M. and A.I.B. and refined with help from X.M. and L.A.B.O. Suggestions and advices for the experiment and data analysis were provided by L.F.C., H.K. and M.G.B. All the authors discussed the results and commented on the manuscript which was written by S.K. and J.W.A.R.

**Competing interests:** The authors declare that they have no competing interests.

**REFERENCES**

1. F. Hübler, M. J. Wolf, D. Beckmann, H. V Löhneysen, Long-range spin-polarized quasiparticle transport in mesoscopic Al superconductors with a Zeeman splitting. *Phys. Rev. Lett.* **109**, 207001 (2012).
2. C. H. L. Quay, D. Chevallier, C. Bena, M. Aprili, Spin imbalance and spin-charge separation in a mesoscopic superconductor. *Nat. Phys.* **9**, 84–88 (2013).
3. J. Linder, J. W. A. Robinson, Superconducting spintronics. *Nat. Phys.* **11**, 307–315 (2015).
4. M. Eschrig, Spin-polarized supercurrents for spintronics. *Phys. Today*. **64**, 43–49 (2011).

5. K. Ohnishi, S. Komori, G. Yang, K. R. Jeon, L. A. B. Olde Olthof, X. Montiel, M. G. Blamire, J. W. A. Robinson, Spin-transport in superconductors. *Appl. Phys. Lett.* **116**, 130501 (2020).
6. F. S. Bergeret, A. F. Volkov, K. B. Efetov, Odd triplet superconductivity and related phenomena in superconductor-ferromagnet structures. *Rev. Mod. Phys.* **77**, 1321–1373 (2005).
7. M. G. Blamire, J. W. A. Robinson, The interface between superconductivity and magnetism : understanding and device prospects. *J. Phys. Condens. Matter*. **26**, 453201 (2014).
8. F. S. Bergeret, A. F. Volkov, K. B. Efetov, Enhancement of the Josephson current by an exchange field in superconductor-ferromagnet structures. *Phys. Rev. Lett.* **86**, 3140–3143 (2001).
9. M. Houzet, A. I. Buzdin, Long range triplet Josephson effect through a ferromagnetic trilayer. *Phys. Rev. B*. **76**, 060504R (2007).
10. S. Mironov, A. Buzdin, Triplet proximity effect in superconducting heterostructures with a half-metallic layer. *Phys. Rev. B*. **92**, 184506 (2015).
11. M. Eschrig, T. L. Ofwander, Triplet supercurrents in clean and disordered half-metallic ferromagnets. *Nat. Phys.* **4**, 138–143 (2008).
12. M. Eschrig, Spin-polarized supercurrents for spintronics : a review of current progress. *Reports Prog. Phys.* **78**, 104501 (2015).
13. T. S. Khaire, M. A. Khasawneh, W. P. Pratt, N. O. Birge, Observation of spin-triplet superconductivity in Co-based Josephson junctions. *Phys. Rev. Lett.* **104**, 137002 (2010).
14. C. Klose, T. S. Khaire, Y. Wang, W. P. Pratt, N. O. Birge, Optimization of spin-triplet supercurrent in ferromagnetic Josephson junctions. *Phys. Rev. Lett.* **108**, 127002 (2012).
15. J. W. A. Robinson, N. Banerjee, M. G. Blamire, Triplet pair correlations and nonmonotonic supercurrent decay with Cr thickness in Nb / Cr / Fe / Nb Josephson devices. *Phys. Rev. B*. **89**, 104505 (2014).
16. Y. Ishikawa, R. Tournibr, J. Filippi, Magnetic properties of Cr-rich Fe-Cr alloys at low temperatures. *J. Phys. Chem. Solids*. **26**, 1727–1745 (1965).
17. J. O. Strom-Olsen, D. F. Wilford, S. K. Burke, B. D. Rainford, The coexistence of spin density wave ordering and spin glass ordering in chromium alloys containing iron. *J. Phys. F Met. Phys.* **9**, L95 (1979).
18. B. Babic, F. Kajzar, G. Parette, Magnetic properties and magnetic interactions in chromium-rich Cr-Fe alloys. *J. Phys. Chem. Solids*. **41**, 1303–1309 (1980).
19. K. Jeon, C. Ciccarelli, A. J. Ferguson, H. Kurebayashi, L. F. Cohen, X. Montiel, M. Eschrig, J. W. A. Robinson, M. G. Blamire, Enhanced spin pumping into superconductors provides evidence for superconducting pure spin currents. *Nat. Mater.* **17**, 499–503 (2018).
20. K. Jeon, C. Ciccarelli, H. Kurebayashi, L. F. Cohen, X. Montiel, M. Eschrig, S. Komori, J. W. A. Robinson, M. G. Blamire, Exchange-field enhancement of superconducting spin pumping. *Phys. Rev. B*. **99**, 024507 (2019).
21. J. Linder, T. Yokoyama, Supercurrent-induced magnetization dynamics in a Josephson junction with two misaligned ferromagnetic layers. *Phys. Rev. B*. **83**, 012501 (2011).
22. B. Baek, W. H. Rippard, S. P. Benz, S. E. Russek, P. D. Dresselhaus, Hybrid superconducting-magnetic memory device using competing order parameters. *Nat. Commun.* **5**, 3888 (2014).

23. A. K. Feofanov, V. A. Oboznov, V. V. Bol'Ginov, J. Lisenfeld, S. Poletto, V. V. Ryazanov, A. N. Rossolenko, M. Khabipov, D. Balashov, A. B. Zorin, P. N. Dmitriev, V. P. Koshelets, A. V. Ustinov, Implementation of superconductor/ferromagnet/superconductor π-shifters in superconducting digital and quantum circuits. *Nat. Phys.* **6**, 593–597 (2010).

24. F. S. Bergeret, I. V. Tokatly, Spin-orbit coupling as a source of long-range triplet proximity effect in superconductor-ferromagnet hybrid structures. *Phys. Rev. B*. **89**, 134517 (2014).

25. X. Montiel, M. Eschrig, Generation of pure superconducting spin current in magnetic heterostructures via nonlocally induced magnetism due to Landau Fermi liquid effects. *Phys. Rev. B*. **98**, 104513 (2018).

26. J. W. A. Robinson, J. D. S. Witt, M. G. Blamire, Controlled injection of spin-triplet supercurrents into a strong ferromagnet. *Science*. **329**, 59–62 (2010).

27. R. S. Keizer, S. T. B. Goennenwein, T. M. Klapwijk, G. Miao, G. Xiao, A. Gupta, A spin triplet supercurrent through the half-metallic ferromagnet $CrO_2$. *Nature*. **439**, 825–827 (2006).

28. M. S. Anwar, M. Veldhorst, A. Brinkman, J. Aarts, M. S. Anwar, M. Veldhorst, A. Brinkman, J. Aarts, Long range supercurrents in ferromagnetic $CrO_2$ using a multilayer contact structure. *Appl. Phys. Lett.* **100**, 052602 (2012).

29. T. Wakamura, N. Hasegawa, K. Ohnishi, Y. Niimi, Y. Otani, Spin Injection into a superconductor with strong spin-orbit coupling. *Phys. Rev. Lett.* **112**, 036602 (2014).

30. M. Morota, Y. Niimi, K. Ohnishi, D. H. Wei, T. Tanaka, H. Kontani, T. Kimura, Y. Otani, Indication of intrinsic spin Hall effect in 4$d$ and 5$d$ transition metals. *Phys. Rev. B*. **83**, 174405 (2011).

31. S. Maekawa, S. O. Valenzuela, E. Saito, T. Kimura, *Spin Current* (Oxford University Press, 2012).

32. C. Bell, G. Burnell, D.-J. Kang, R. H. Hadfield, M. J. Kappers, M. G. Blamire, Fabrication of nanoscale heterostructure devices with a focused ion beam. *Nanotechnology*. **14**, 630–632 (2003).

33. M. Hurd, G. Wendin, Superconducting current in a ballistic double superconducting—normal-metal—superconducting structure. *Phys. Rev. B*. **51**, 3754–3759 (1995).

34. V. C. Y. Chang, C. S. Chu, Andreev-level tunneling in a ballistic double superconductor – normal-metal – superconductor junction. *Phys. Rev. B*. **55**, 6004–6014 (1997).

35. J. A. Ouassou, J. Linder, Spin-switch Josephson junctions with magnetically tunable sin (δφ/n) current-phase relation. *Phys. Rev. B*. **96**, 064516 (2017).

36. H. Zhang, J. W. Lynn, C. F. Majkrzak, S. K. Satija, J. H. Kang, X. D. Wu, Measurements of magnetic screening lengths in superconducting Nb thin films by polarized neutron refiectometry. *Appl. Phys. Lett.* **52**, 10395–10404 (1995).

37. E. Nazaretski, J. P. Thibodaux, I. Vekhter, L. Civale, J. D. Thompson, R. Movshovich, Direct measurements of the penetration depth in a superconducting film using magnetic force microscopy. *Appl. Phys. Lett.* **95**, 262502 (2009).

38. N. W. Ascroft, N. D. Mermin, *Solid State Physics* (Holt, Rinehart and Winston, New York, 1976).

39. M. G. Blamire, C. B. Smiet, N. Banerjee, J. W. A. Robinson, Field modulation of the critical current in magnetic Josephson junctions. *Supercond. Sci. Technol.* **26**, 055017 (2013).

40. Y. Kalcheim, O. Millo, A. Di Bernardo, A. Pal, J. W. A. Robinson, Inverse proximity effect at superconductor-ferromagnet interfaces: Evidence for induced triplet pairing in the superconductor. *Phys. Rev. B*. **92**, 060501R (2015).
41. J. Linder, A. Sudbø, T. Yokoyama, R. Grein, M. Eschrig, Signature of odd-frequency pairing correlations induced by a magnetic interface. *Phys. Rev. B*. **81**, 214504 (2010).
42. D. Bozec, thesis, Leeds University, Leeds, West Yorkshire, England (2000).
43. C. Richard, A. Buzdin, M. Houzet, J. S. Meyer, Signatures of odd-frequency correlations in the Josephson current of superconductor/ferromagnet hybrid junctions. *Phys. Rev. B*. **92**, 094509 (2015).

# Supplementary Information

## 1. Spin-transport in normal state Nb

Figure S1 shows the electrical resistance ($R$) versus an in-plane magnetic field ($H$) for Nb(300 nm)/Cr(1 nm)/Fe(4.8 nm)/Cu(10 nm)/Fe(2.4 nm)/Cr(1 nm)/Nb(300 nm) (left-axis) and Nb(300 nm)/Cr(1 nm)/Fe(4.8 nm)/Nb(4.8 nm)/Fe(2.4 nm)/Cr(1 nm)/Nb(300 nm) (right-axis) devices at 10 K. A mismatch between the coercive fields of the 4.8-nm-thick Fe layer and the 2.4-nm-thick Fe layer leads to an increase in $R$ at $\mu_0 H \approx 50$ mT due to giant magnetoresistance (GMR) effect in the device with a 10-nm-thick Cu spacer but not in the device with a 4.8-nm-thick Nb spacer. The current is applied perpendicular to the plane in the device and hence the thickness of the spacer should be less than the spin-diffusion length to observe GMR. The absence of GMR in the device with a Nb spacer indicates a short spin-diffusion length (< 5 nm) in the normal state of Nb in these devices which agrees with the decay envelope of triplet supercurrents in Fig. 3A in the main text.

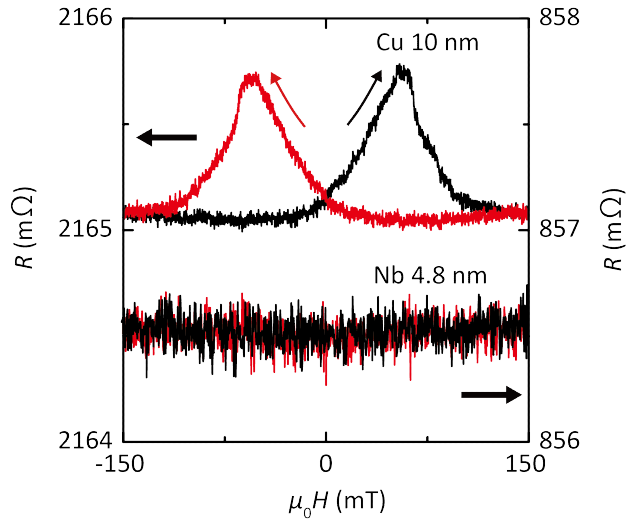

**Fig. S1.** $R$ ($H$) for Nb(300)/Cr(1)/Fe(4.8)/Cu(10)/Fe(2.4)/Cr(1)/Nb(300) (left-axis) and Nb(300)/Cr(1)/Fe(4.8)/Nb(4.8)/Fe(2.4)/Cr(1)/Nb(300) (right-axis) devices at $T$ = 10 K.

## 2. Theory of the suppression of spin-triplet Josephson currents in a singlet superconductor

We consider a $S_1/F_1/S'/F_2/S_2$ Josephson junction (see Fig. S2) consisting of atomically thin superconductors ($S_1$ and $S_2$), ferromagnets ($F_1$ and $F_2$) and a central superconductor ($S'$). The neighbouring layers are coupled by the transfer integrals $t_i$, ($i$ = 1, 2, 3, 4) of the tight-binding model. The critical temperature of the superconducting leads $S_1$ and $S_2$ ($T_{c1} = T_{c2}$) is higher than that of the central superconductor $S'$ ($T_{c0}$). Thus, the central layer $S'$ can be both in the normal and in the superconducting states at $T < T_{c1}$. We assume that $T \approx T_{c0} < T_{c1}$, $t_i \ll T_{c0}$ and the interlayer tunneling conserves the momentum. Also, we assume that the $S_1/F_1$ and $F_2/S_2$ interfaces are magnetized. The misalignment angle $\theta_i$ ($i$ = 1, 2) between the exchange field $\mathbf{h}_i = h_i(\cos\theta_i \, \mathbf{z} + \sin\theta_i \, \mathbf{x})$ at the $S_i/F_i$ interface and the spin-rotation axis $\mathbf{z}$ in the $F_i$ layer gives rise to the emergence of the spin-triplet superconducting correlations. We assume 100 % spin polarization of $F_1$ and $F_2$ layers and therefore the transport of minority spin triplet pairs is blocked.

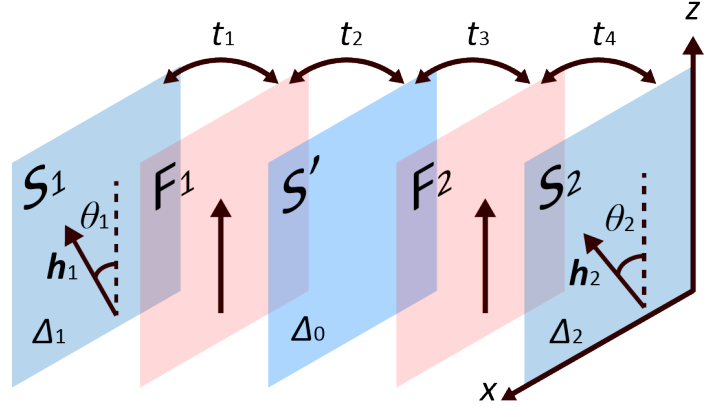

**Fig. S2. S$_1$/F$_1$/S'/F$_2$/S$_2$ Josephson junction consisting of atomically thin layers.**

The superconducting gaps in S', S$_1$ and S$_2$ layers are $\Delta_0$, $\Delta_1 = |\Delta_1|e^{-i\varphi/2}$ and $\Delta_2 = |\Delta_2|e^{i\varphi/2}$ with $|\Delta_1| = |\Delta_2|$ (the phase difference across the junction equals to $\varphi$). The energy spectrum in the superconductors is $\xi(\mathbf{p})$, while in the ferromagnets it is spin-dependent: $\xi_\uparrow = \xi(\mathbf{p})$ and $\xi_\downarrow = +\infty$. We denote the electron annihilation operators in S$_1$, F$_1$, S', F$_2$, S$_2$ layers as $\hat{\eta}$, $\hat{\psi}$, $\hat{\phi}$, $\hat{\tilde{\psi}}$ and $\hat{\tilde{\eta}}$. The Hamiltonian in the system under consideration is

$$\hat{H} = \hat{H}_0 + \hat{H}_{\text{BCS}} + \hat{H}_t, \tag{S1}$$

with $\hat{H}_0$ the single particle Hamiltonian describing the kinetic energy in the five layers. $\hat{H}_0$ writes:

$$\hat{H}_0 = \sum_{\mathbf{p};\alpha,\beta=\uparrow,\downarrow} \xi(\mathbf{p})\,\hat{\phi}^\dagger_{\mathbf{p},\alpha}\hat{\phi}_{\mathbf{p},\beta}\delta_{\alpha\beta} + \hat{P}_{\alpha\beta}\hat{\psi}^\dagger_{\mathbf{p},\alpha}\hat{\psi}_{\mathbf{p},\beta} + \hat{P}_{\alpha\beta}\hat{\tilde{\psi}}^\dagger_{\mathbf{p},\alpha}\hat{\tilde{\psi}}_{\mathbf{p},\beta} + \hat{C}^{(1)}_{\alpha\beta}\hat{\eta}^\dagger_{\mathbf{p},\alpha}\hat{\eta}_{\mathbf{p},\beta} + \hat{C}^{(2)}_{\alpha\beta}\hat{\tilde{\eta}}^\dagger_{\mathbf{p},\alpha}\hat{\tilde{\eta}}_{\mathbf{p},\beta} \tag{S2}$$

where the coefficient $\hat{C}_i$ ($i$ = 1,2) describes the modified kinetic energy due to the magnetized interfaces and $\hat{P}$ is the spin-dependent kinetic energy of the ferromagnets (assumed identical for F$_1$ and F$_2$), given by

$$\hat{C}^{(i)} = \begin{pmatrix} \xi - h_i\cos\theta_i & -h_i\sin\theta_i \\ -h_i\sin\theta_i & \xi + h_i\cos\theta_i \end{pmatrix}, \qquad \hat{P} = \begin{pmatrix} \xi_\uparrow & 0 \\ 0 & \xi_\downarrow \end{pmatrix} = \begin{pmatrix} \xi & 0 \\ 0 & \infty \end{pmatrix}.$$

The second term in (S1), $\hat{H}_{\text{BCS}}$ describes the superconductivity in the three superconductors as

$$\hat{H}_{\text{BCS}} = \sum_{\mathbf{p}} \Delta_0 \hat{\phi}^\dagger_{\mathbf{p},\uparrow}\hat{\phi}^\dagger_{-\mathbf{p},\downarrow} + \Delta_0^* \hat{\phi}_{-\mathbf{p},\downarrow}\hat{\phi}_{\mathbf{p},\uparrow} + \Delta_1 \hat{\eta}^\dagger_{\mathbf{p},\uparrow}\hat{\eta}^\dagger_{-\mathbf{p},\downarrow} + \Delta_1^* \hat{\eta}_{-\mathbf{p},\downarrow}\hat{\eta}_{\mathbf{p},\uparrow} + \Delta_2 \hat{\tilde{\eta}}^\dagger_{\mathbf{p},\uparrow}\hat{\tilde{\eta}}^\dagger_{-\mathbf{p},\downarrow} + \Delta_2^* \hat{\tilde{\eta}}_{-\mathbf{p},\downarrow}\hat{\tilde{\eta}}_{\mathbf{p},\uparrow} \tag{S3}$$

and $\hat{H}_t$ is the tunnelling Hamiltonian given by

$$\widehat{H}_t = \sum_{\mathbf{p};\alpha} t_1 \left( \eta^\dagger_{\mathbf{p},\alpha} \psi_{\mathbf{p},\alpha} + \psi^\dagger_{\mathbf{p},\alpha} \eta_{\mathbf{p},\alpha} \right) + t_2 (\psi^\dagger_{\mathbf{p},\alpha} \phi_{\mathbf{p},\alpha} + \phi^\dagger_{\mathbf{p},\alpha} \psi_{\mathbf{p},\alpha}) + t_3 (\phi^\dagger_{\mathbf{p},\alpha} \tilde{\psi}_{\mathbf{p},\alpha} + \tilde{\psi}^\dagger_{\mathbf{p},\alpha} \phi_{\mathbf{p},\alpha})$$
$$+ t_4 \left( \tilde{\psi}^\dagger_{\mathbf{p},\alpha} \tilde{\eta}_{\mathbf{p},\alpha} + \tilde{\eta}^\dagger_{\mathbf{p},\alpha} \tilde{\psi}_{\mathbf{p},\alpha} \right). \tag{S4}$$

The commutation relations with the Hamiltonian are

$$[\widehat{H}, \hat{\phi}_{\mathbf{p},\uparrow}] = -\xi \hat{\phi}_{\mathbf{p},\uparrow} + \Delta_0 \hat{\phi}^\dagger_{-\mathbf{p},\downarrow} - t_2 \hat{\psi}_{\mathbf{p},\uparrow} - t_3 \hat{\tilde{\psi}}_{\mathbf{p},\uparrow}, \qquad [\widehat{H}, \hat{\phi}^\dagger_{-\mathbf{p},\downarrow}] = \xi \hat{\phi}^\dagger_{-\mathbf{p},\downarrow} + \Delta_0^* \hat{\phi}_{\mathbf{p},\uparrow} + t_2 \hat{\psi}^\dagger_{-\mathbf{p},\downarrow} + t_3 \hat{\tilde{\psi}}^\dagger_{-\mathbf{p},\downarrow},$$

$$[\widehat{H}, \hat{\psi}_{\mathbf{p},\uparrow}] = -\sum_\beta P_{\uparrow\beta} \hat{\psi}_{\mathbf{p},\beta} - t_1 \hat{\eta}_{\mathbf{p},\uparrow} - t_2 \hat{\phi}_{\mathbf{p},\uparrow}, \qquad [\widehat{H}, \hat{\psi}^\dagger_{-\mathbf{p},\downarrow}] = \sum_\alpha P_{\alpha\downarrow} \hat{\psi}^\dagger_{-\mathbf{p},\alpha} + t_1 \hat{\eta}^\dagger_{-\mathbf{p},\downarrow} + t_2 \hat{\phi}^\dagger_{-\mathbf{p},\downarrow},$$

$$[\widehat{H}, \hat{\tilde{\psi}}_{\mathbf{p},\uparrow}] = -\sum_\beta P_{\uparrow\beta} \hat{\tilde{\psi}}_{\mathbf{p},\beta} - t_3 \hat{\phi}_{\mathbf{p},\uparrow} - t_4 \hat{\tilde{\eta}}_{\mathbf{p},\uparrow}, \qquad [\widehat{H}, \hat{\tilde{\psi}}^\dagger_{-\mathbf{p},\downarrow}] = \sum_\alpha P_{\alpha\downarrow} \hat{\tilde{\psi}}^\dagger_{-\mathbf{p},\alpha} + t_3 \hat{\phi}^\dagger_{-\mathbf{p},\downarrow} + t_4 \hat{\tilde{\eta}}^\dagger_{-\mathbf{p},\downarrow},$$

$$[\widehat{H}, \hat{\eta}_{\mathbf{p},\uparrow}] = -\sum_\beta \hat{C}^{(1)}_{\uparrow\beta} \hat{\eta}_{\mathbf{p},\beta} + \Delta_1 \hat{\eta}^\dagger_{-\mathbf{p},\downarrow} - t_1 \hat{\psi}_{\mathbf{p},\uparrow}, \qquad [\widehat{H}, \hat{\eta}^\dagger_{-\mathbf{p},\downarrow}] = \sum_\alpha \hat{C}^{(1)}_{\alpha\downarrow} \hat{\eta}^\dagger_{-\mathbf{p},\alpha} + \Delta_1^* \hat{\eta}_{\mathbf{p},\uparrow} + t_1 \hat{\psi}^\dagger_{-\mathbf{p},\downarrow},$$

$$[\widehat{H}, \hat{\tilde{\eta}}_{\mathbf{p},\uparrow}] = -\sum_\beta \hat{C}^{(2)}_{\uparrow\beta} \hat{\tilde{\eta}}_{\mathbf{p},\beta} + \Delta_2 \hat{\tilde{\eta}}^\dagger_{-\mathbf{p},\downarrow} - t_4 \hat{\tilde{\psi}}_{\mathbf{p},\uparrow}, \qquad [\widehat{H}, \hat{\tilde{\eta}}^\dagger_{-\mathbf{p},\downarrow}] = \sum_\alpha \hat{C}^{(2)}_{\alpha\downarrow} \hat{\tilde{\eta}}^\dagger_{-\mathbf{p},\beta} + \Delta_2^* \hat{\tilde{\eta}}_{\mathbf{p},\uparrow} + t_4 \hat{\tilde{\psi}}^\dagger_{-\mathbf{p},\downarrow}.$$

We assume the coherent electron tunneling between the layers, which preserves the in-plane momentum **p**. Using the Liouville equation

$$i \frac{\partial \Psi}{\partial \tau} = [\widehat{H}, \Psi],$$

we introduce the following Green's functions in the imaginary-time representation:

$$G_{\alpha\beta}(\mathbf{p}; \tau_1, \tau_2) = -\langle T_\tau \hat{\phi}_{\mathbf{p},\alpha}(\tau_1) \hat{\phi}^\dagger_{\mathbf{p},\beta}(\tau_2) \rangle, \qquad F^\dagger_{\alpha\beta}(\mathbf{p}; \tau_1, \tau_2) = \langle T_\tau \hat{\phi}^\dagger_{-\mathbf{p},\alpha}(\tau_1) \hat{\phi}^\dagger_{\mathbf{p},\beta}(\tau_2) \rangle,$$

$$E^\psi_{\alpha\beta}(\mathbf{p}; \tau_1, \tau_2) = -\langle T_\tau \hat{\psi}_{\mathbf{p},\alpha}(\tau_1) \hat{\phi}^\dagger_{\mathbf{p},\beta}(\tau_2) \rangle, \qquad F^{\psi\dagger}_{\alpha\beta}(\mathbf{p}; \tau_1, \tau_2) = \langle T_\tau \hat{\psi}^\dagger_{-\mathbf{p},\alpha}(\tau_1) \hat{\phi}^\dagger_{\mathbf{p},\beta}(\tau_2) \rangle,$$

$$E^{\tilde{\psi}}_{\alpha\beta}(\mathbf{p}; \tau_1, \tau_2) = -\langle T_\tau \hat{\tilde{\psi}}_{\mathbf{p},\alpha}(\tau_1) \hat{\phi}^\dagger_{\mathbf{p},\beta}(\tau_2) \rangle, \qquad F^{\tilde{\psi}\dagger}_{\alpha\beta}(\mathbf{p}; \tau_1, \tau_2) = \langle T_\tau \hat{\tilde{\psi}}^\dagger_{-\mathbf{p},\alpha}(\tau_1) \hat{\phi}^\dagger_{\mathbf{p},\beta}(\tau_2) \rangle,$$

$$E^\eta_{\alpha\beta}(\mathbf{p}; \tau_1, \tau_2) = -\langle T_\tau \hat{\eta}_{\mathbf{p},\alpha}(\tau_1) \hat{\phi}^\dagger_{\mathbf{p},\beta}(\tau_2) \rangle, \qquad F^{\eta\dagger}_{\alpha\beta}(\mathbf{p}; \tau_1, \tau_2) = \langle T_\tau \hat{\eta}^\dagger_{-\mathbf{p},\alpha}(\tau_1) \hat{\phi}^\dagger_{\mathbf{p},\beta}(\tau_2) \rangle,$$

$$E^{\tilde{\eta}}_{\alpha\beta}(\mathbf{p}; \tau_1, \tau_2) = -\langle T_\tau \hat{\tilde{\eta}}_{\mathbf{p},\alpha}(\tau_1) \hat{\phi}^\dagger_{\mathbf{p},\beta}(\tau_2) \rangle, \qquad F^{\tilde{\eta}\dagger}_{\alpha\beta}(\mathbf{p}; \tau_1, \tau_2) = \langle T_\tau \hat{\tilde{\eta}}^\dagger_{-\mathbf{p},\alpha}(\tau_1) \hat{\phi}^\dagger_{\mathbf{p},\beta}(\tau_2) \rangle,$$

where $G$ and $F^\dagger$ are the single-particle and anomalous Green's function in S', respectively. The tunneling Green's functions $E^\psi$, $E^{\tilde{\psi}}$, $E^\eta$ and $E^{\tilde{\eta}}$ represent the tunneling of a particle from S' to F$_1$, F$_2$, S$_1$ and S$_2$, respectively. Finally, the anomalous Green's functions $F^{\psi\dagger}$, $F^{\tilde{\psi}\dagger}$, $F^{\eta\dagger}$ and $F^{\tilde{\eta}\dagger}$ are associated with the creation of a Cooper pair in which one electron is located in S' and the other electron is in F$_1$, F$_2$, S$_1$ and S$_2$, respectively.

Rewriting the commutation relations in terms of the Green's functions and applying the Fourier transform such that $i\partial\Psi/\partial\tau = -i\omega\Psi$, we find the following closed set of matrices Gor'kov equations in frequency representation:

$$\text{for S':} \quad \begin{cases} (i\omega - \xi)G + i\Delta_0\sigma_y F^\dagger - t_2 IE^\psi - t_3 IE^{\tilde{\psi}} = I, \\ (i\omega + \xi)F^\dagger - i\Delta_0^*\sigma_y G + t_2 IF^{\psi\dagger} + t_3 IF^{\tilde{\psi}\dagger} = 0, \end{cases} \quad (S5)$$

$$\text{for F}_1\text{:} \quad \begin{cases} (i\omega - \hat{P})E^\psi - t_1 IE^\eta - t_2 IG = 0, \\ (i\omega + \hat{P})F^{\psi\dagger} + t_1 IF^{\eta\dagger} + t_2 IF^\dagger = 0, \end{cases} \quad (S6)$$

$$\text{for F}_2\text{:} \quad \begin{cases} (i\omega - \hat{P})E^{\tilde{\psi}} - t_3 IG - t_4 IE^{\tilde{\eta}} = 0, \\ (i\omega + \hat{P})F^{\tilde{\psi}\dagger} + t_3 IF^\dagger + t_4 IF^{\tilde{\eta}\dagger} = 0, \end{cases} \quad (S7)$$

$$\text{for S}_1\text{:} \quad \begin{cases} (i\omega - \hat{C}^{(1)})E^\eta + i\Delta_1\sigma_y F^{\eta\dagger} - t_1 IE^\psi = 0, \\ (i\omega + \hat{C}^{(1)})F^{\eta\dagger} - i\Delta_1^*\sigma_y E^\eta + t_1 IF^{\psi\dagger} = 0, \end{cases} \quad (S8)$$

$$\text{for S}_2\text{:} \quad \begin{cases} (i\omega - \hat{C}^{(2)})E^{\tilde{\eta}} + i\Delta_2\sigma_y F^{\tilde{\eta}\dagger} - t_4 IE^{\tilde{\psi}} = 0, \\ (i\omega + \hat{C}^{(2)})F^{\tilde{\eta}\dagger} - i\Delta_2^*\sigma_y E^{\tilde{\eta}} + t_4 IF^{\tilde{\psi}\dagger} = 0. \end{cases} \quad (S9)$$

where $I$ is the 2 × 2 identity matrix and $\sigma_y$ is the second Pauli matrix.

The definition of the current through the junction is imposed by the tunnelling Hamiltonian model. In this description, the charge current corresponds to the number of particles travelling from one layer to another, as described by the tunnelling Green's function $E$. Since the charge current flows across all the layers, it can be obtained from the Green's function at an arbitrary layer. The model presented here focuses on the Green's functions in the central S' layer, such that we express the current in terms of $E^\psi$. The current consists of the sum of spin-up and spin-down currents, however, since we consider fully polarized ferromagnets, we only need to take the $E^\psi_{\alpha\alpha}$ component into account.

Hence, the Josephson current density ($j_y$) across the junction is expressed via the Fourier component $E^\psi_{\alpha\alpha}(\mathbf{p};\omega)$ of the off-diagonal Matsubara Green function $E^\psi_{\alpha\beta}(\mathbf{p};\tau_1,\tau_2) = -\langle T_\tau \hat{\psi}_{\mathbf{p},\alpha}(\tau_1)\hat{\phi}^\dagger_{\mathbf{p},\beta}(\tau_2)\rangle$:

$$j_y = -2ev_0 t_2 T \text{Im} \sum_{\omega=-\infty}^{\infty} \int_{-\infty}^{\infty} E^\psi_{\alpha\alpha}(\mathbf{p};\omega)\, d\xi, \quad (S10)$$

where $v_0$ and $T_\tau$ are the electron density of states at the Fermi level and the time-ordered product for the imaginary time $\tau$, respectively.

By solving the system above, we find the exact expression for $E^\psi$. We obtain $E^\psi_{\downarrow\downarrow}(\mathbf{p};\omega) = 0$, which corresponds to the absence of the spin-down state, resulting from fully spin-polarized ferromagnets. By expanding $E^{\psi}_{\uparrow\uparrow}$ up to seventh order over $t_i \ll T$, ($i = 1, 2, 3, 4$) (assuming $h_2, h_1 \ll T \approx T_{c1}$ and $|\Delta_0| \ll T$) we find:

$$\text{Im} \sum_{\omega=-\infty}^{\infty} \int_{-\infty}^{\infty} E^{\psi}_{\uparrow\uparrow}(\mathbf{p}; \omega) d\xi = 4t_1^2 t_2^2 t_3^2 t_4^2 \text{Im}[\Delta_1 \Delta_2^*](h_1 \sin\theta_1)(h_2 \sin\theta_2)(a - b|\Delta_0|^2),$$

where the coefficients are given by

$$a = -\sum_{\omega=-\infty}^{\infty} \int_{-\infty}^{\infty} \frac{\omega^2}{(i\omega - \xi)^4 (i\omega - \xi)^2 [(i\omega + \xi)(i\omega - \xi) - |\Delta_1|^2]^4} d\xi$$

$$= \sum_{\omega>0} \frac{\pi \omega^2}{8|\Delta_1|^{14}} \left[ -640\omega - \frac{16|\Delta_1|^4}{\omega^3} + \frac{4|\Delta_1|^6}{\omega^5} \right.$$

$$\left. + \frac{640\omega^8 + 2240\omega^6 |\Delta_1|^2 + 2816\omega^4 |\Delta_1|^4 + 1452\omega^2 |\Delta_1|^6 + 231|\Delta_1|^8}{(\omega^2 + |\Delta_1|^2)^{7/2}} \right],$$

$$b = \sum_{\omega=-\infty}^{\infty} \int_{-\infty}^{\infty} \frac{2\omega^2}{(i\omega + \xi)^3 (i\omega - \xi)^5 [(i\omega + \xi)(i\omega - \xi) - |\Delta_1|^2]^4} d\xi$$

$$= \sum_{\omega>0} \frac{\pi \omega^2}{16|\Delta_1|^{16}} \left[ 4480\omega + \frac{160|\Delta_1|^4}{\omega^3} - \frac{64|\Delta_1|^6}{\omega^5} + \frac{15|\Delta_1|^8}{\omega^7} \right.$$

$$\left. - \frac{4(1128\omega^8 + 3920\omega^6 |\Delta_1|^2 + 4940\omega^4 |\Delta_1|^4 + 2574\omega^2 |\Delta_1|^6 + 429|\Delta_1|^8)}{(\omega^2 + |\Delta_1|^2)^{7/2}} \right].$$

such that the Josephson current density (S10) becomes

$$j_y = 8ev_0 t_1^2 t_2^2 t_3^2 t_4^2 |\Delta_1|^2 T (h_1 \sin\theta_1)(h_2 \sin\theta_2)(a - b|\Delta_0|^2)\sin\varphi. \tag{S11}$$

Note that the temperature dependence of the critical current in the absence of the singlet superconductivity ($|\Delta_0| = 0$) may be non-monotonous, similar to the results obtained by Eschrig and Löfwander[1,2] for the triplet supercurrents in a half-metallic Josephson junction with spin-active interfaces.

At low temperatures, $T \ll T_{c2}$ and $|\Delta_1| \gg T$, the coefficients $a$ and $b$ reduce to

$$a \approx \frac{7\zeta(3)}{16\pi^2 |\Delta_1|^8 T^3}, \qquad b \approx \frac{465\zeta(5)}{512\pi^4 |\Delta_1|^8 T^5}.$$

where $\zeta$ is the Riemann zeta function.

At a temperature of $T \approx T_{c0}$, the condition above is satisfied when $|\Delta_1(T_{c0})| \gg T_{c0}$ (i.e. $T_{c1} \gg T_{c0}$). In this case, the Josephson current density is

$$j_y = \frac{56\zeta(3)ev_0 t_1^2 t_2^2 t_3^2 t_4^2 h_1 h_2 \sin\theta_1 \sin\theta_2}{16\pi^2 |\Delta_1|^6 T^2}\left(1 - \frac{465\zeta(5)}{224\pi^2\zeta(3)} \frac{|\Delta_0|^2}{T^2}\right)\sin\varphi. \tag{S12}$$

The divergence of the critical current in eq. (S12) at low temperature is related to the expansion over $t$ and should be cut off at $T \approx t$. This shows that the singlet superconductivity suppresses the triplet Josephson current. Superconductivity in the central S' layer suppresses the triplet component of the anomalous Green's function $F^\dagger$. The fourth term in the expansion of $F^\dagger$ over $t$ is

$$F_{11}^\dagger(\mathbf{p};\omega) = \frac{\alpha_1 t_1^2 t_2^2 \Delta_1^* + \alpha_2 t_3^2 t_4^2 \Delta_2^*}{(i\omega - \xi)(i\omega + \xi)}\left(1 + \frac{|\Delta_0|^2}{\omega^2 + \xi^2}\right)^{-2}, \tag{S13}$$

with coefficients $\alpha_{1,2} = d_{1,2}/\{(i\omega - \xi)(i\omega + \xi)\}$, $d_{1,2} = \left[\hat{A}_{1,2}\hat{I}\{i\omega - \hat{C}^{(1,2)}\}^{-1}\right]_{11}$ and $\hat{A}_{1,2} = \left[\{i\omega + \hat{C}^{(1,2)}\} + |\Delta_{1,2}|^2 \hat{I}\{i\omega - \hat{C}^{(1,2)}\}^{-1}\hat{I}\right]^{-1}$.

We conclude that the suppression of the triplet component by the singlet superconducting correlations in the central S' layer results in damping of the Josephson current through a S$_1$/F$_1$/S'/F$_2$/S$_2$ Josephson junction. In the case of a symmetric junctions with $t_1 = t_2 = t_3 = t_4 = t$ and $h_1 = h_2 = h$, for temperature $T_{c1} \gg T_{c0}$ with $T \approx T_{c0}$, the Josephson current is

$$j_y = j_0 \left(\frac{t}{T_{c0}}\right)^8 \left(\frac{h}{T_{c0}}\right)^2 \left(\frac{T_{c0}}{|\Delta_1|}\right)^6 \left(\frac{T_{c0}}{T}\right)^2 \left(1 - \beta \frac{|\Delta_0|^2}{T^2}\right)\sin\theta_1 \sin\theta_2 \sin\varphi, \tag{S14}$$

with $j_0 = 56\zeta(3)ev_0 T_{c0}^2/(16\pi^2)$ and $\beta = 465\zeta(5)/\{224\pi^2\zeta(3)\} \approx 0.2$.

**Supplementary References**

1. M. Eschrig, T. L. Ofwander, Triplet supercurrents in clean and disordered half-metallic ferromagnets. *Nat. Phys.* **4**, 138–143 (2008).
2. M. Eschrig, Spin-polarized supercurrents for spintronics: a review of current progress. *Reports Prog. Phys.* **78**, 104501 (2015).